\documentclass[aip,cha,reprint]{revtex4-1}

\usepackage{graphicx}  \usepackage[english]{babel}
\usepackage{pifont}
\usepackage{amssymb, amsfonts, amsmath}  \usepackage{units}
\usepackage{lineno}
\usepackage{hyperref}

\DeclareMathOperator{\sgn}{sgn}

\begin{document}

\setcounter{page}{0}
\textsf{\textbf{
{\onecolumngrid \noindent \Large
Copyright 2014 American Institute of Physics. This article may be downloaded for personal use only. Any other use requires prior permission of the author and the American Institute of Physics.\vspace{1\baselineskip}\\
The following article appeared in Porz S, Kiel M, Lehnertz K, Chaos 24, 033112 (2014) and may be found at \url{http://dx.doi.org/10.1063/1.4890568}.
}
}}
\newpage

\title{
    Can spurious indications for phase synchronization due to superimposed signals be avoided?
}

\date{\today}

\author{Stephan \surname{Porz}}
\email{s.porz@uni-bonn.de}
\affiliation{Department of Epileptology, University of Bonn, Sigmund-Freud-Str. 25, 53105 Bonn, Germany}
\affiliation{Helmholtz Institute for Radiation and Nuclear Physics, University of Bonn, Nussallee 14--16, 53115 Bonn, Germany}

\author{Matth\"aus \surname{Kiel}}
\email{matthaeus.kiel@kit.edu}
\affiliation{Department of Epileptology, University of Bonn, Sigmund-Freud-Str. 25, 53105 Bonn, Germany}
\affiliation{Helmholtz Institute for Radiation and Nuclear Physics, University of Bonn, Nussallee 14--16, 53115 Bonn, Germany}
\affiliation{Institute for Meteorology and Climate Research -- Atmospheric Trace Gases and Remote Sensing, Karlsruhe Institute of Technology, H.-v.-Helmholtz-Platz 1, 76344 Leopoldshafen, Germany}

\author{Klaus \surname{Lehnertz}}
\email{klaus.lehnertz@ukb.uni-bonn.de}
\affiliation{Department of Epileptology, University of Bonn, Sigmund-Freud-Str. 25, 53105 Bonn, Germany}
\affiliation{Helmholtz Institute for Radiation and Nuclear Physics, University of Bonn, Nussallee 14--16, 53115 Bonn, Germany}
\affiliation{Interdisciplinary Center for Complex Systems, University of Bonn, Br\"uhler Str. 164, 53175 Bonn, Germany}

\begin{abstract}
We investigate the relative merit of phase-based methods---mean phase coherence, unweighted and weighted phase lag index---for estimating the strength of interactions between dynamical systems from empirical time series which are affected by common sources and noise.
By numerically analyzing the interaction dynamics of coupled model systems, we compare these methods to each other with respect to their ability to distinguish between different levels of coupling for various simulated experimental situations.
We complement our numerical studies by investigating consistency and temporal variations of the strength of interactions within and between brain regions using intracranial electroencephalographic recordings from an epilepsy patient.
Our findings indicate that the unweighted and weighted phase lag index are less prone to the influence of common sources but that this advantage may lead to constrictions limiting the applicability of these methods.
\end{abstract}

\maketitle

\begin{quotation}
The study of synchronization phenomena in coupled dynamical systems is an active field of research in many scientific disciplines including empirical studies on the dynamics on and of complex networks.
A number of time series analysis techniques are available that allow one to capture both linear and nonlinear aspects of interactions.
In many experimental situations, however, spurious indications of interactions may arise due to the presence of so-called common sources, that are caused, for example, by an overly dense spatial sampling of the systems' dynamics.
In order to avoid severe misinterpretations, phase-based estimators have recently been developed that appear to be immune to common sources.
Here we compare the relative merit of these improved estimators with widely used ones that are also based on phase synchronization.
We show that common sources have a reduced influence on the improved estimators, achieving maximum efficiency for controlled situations.
However, we also observe several unwanted side effects that may reduce their practical value.
\end{quotation}

\section*{Introduction}
Synchronization and related complex interaction phenomena are ubiquitous in nature and play an important role in numerous scientific fields, ranging from physics to the neurosciences\cite{Pikovsky_Book2001, Boccaletti2002, Osipov2007, Lehnertz2009b, FellAxmacher2011, Bloch2013}.
Recently, there is an increasing interest to understand synchronization phenomena in complex networks, as they have been recognized to be powerful representations of spatially extended dynamical systems and can advance our understanding of their dynamics\cite{Arenas2008, Havlin2012, Siegel2012, Lehnertz2014}.
Characterizing the coupling between interacting (sub-)systems as well as deriving weighted and directed networks from empirical data requires estimating the strength and direction of an interaction.
Over the last years, a large number of linear and nonlinear analysis techniques have been proposed that allow a data-driven quantification of these interaction properties\cite{Brillinger1981, Pikovsky_Book2001, Boccaletti2002, Pereda2005, Hlavackova2007, Marwan2007, Lehnertz2009b, Lehnertz2011, Sommerlade2012}.
When analyzing empirical data, however, one is often faced with the problem to decide whether a given value of some estimator indeed indicates the strength or the directionality of an interaction or whether it merely reflects the influence of other factors such as noise, biases, sample size, or statistical issues.
The use of analysis techniques that cannot distinguish between functional interactions between subsystems and spurious interactions (e.g., caused by sampling the same subsystem, i.e., a common source) can lead to severe misinterpretations\cite{Smirnov2003, Guevara2005, Meinecke2005, Schelter2007, Smirnov2009, Tognoli2009, Duggento2012, Smirnov2013b}
that can even affect network properties\cite{Bialonski2010, Bialonski2011b, Palus2011, Peraza2012, Zerenner2014}.

For widely used estimators for phase synchronization\cite{Hoke1988, Lachaux1999, Mormann2000}, modifications and extensions\cite{Stam2007c, Vinck2011} have been proposed that appear to be much less affected by the influence of common sources.
Studies investigating the relative performance of estimators, however, are rare and had been carried out for particular applications only\cite{Peraza2012, Aydore2013, Gordon2013}.
Here we extend these studies and investigate the relative merit of estimators for a data-driven quantification of the strength of interactions.
For this purpose, we create a \emph{controlled} setting for a comparison of estimators, mimicking experimental situations.
Using the dynamics of various paradigmatic model systems, ranging from coupled phase oscillators to coupled structurally identical and non-identical nonlinear oscillators with chaotic dynamics, we estimate the strength of interactions from noisy model time series influenced by common sources.
Complementing these numerical studies we then investigate consistency and temporal variations of the strength of interactions within and between brain regions using long-lasting intracranial electroencephalographic recordings from an epilepsy patient.
We show that the modified estimators can help to reduce spurious indications of interactions caused by the influence of common sources, but this capacity may come along with a loss of important spatial and temporal aspects of the interaction dynamics.

\section*{Methods}
\subsection*{Estimating the strength of interactions}
Let $\Phi_{l}(j), l \in \{a, b\}, j = 1, \ldots, N$ denote phase time series from systems $a$ and $b$, where $N$ is the length of the time series.
From field data, phases can be derived, e.g., with the Hilbert, the wavelet, or the Gabor transform\cite{LeVanQuyen2001c, Bruns2004, Palus2005, Wacker2011}.
The mean phase coherence\cite{Hoke1988, Lachaux1999, Mormann2000} is a widely used estimator for the strength of interactions, and is defined as:
\begin{equation}\label{eq:MPC}
R = \left| \frac{1}{N} \sum_{j=1}^{N} \exp \left(i\left(\Phi_{a}(j) - \Phi_{b}(j) \right) \right) \right|.
\end{equation}
$R$ is confined to the interval $[0,1]$, where $R = 1$ indicates fully synchronized systems.
This estimator was shown to be influenced by spurious correlations due to common sources\cite{Stam2007c}.
In this case, it will attain artificially increased values which can lead to misinterpretations regarding the strength of an interaction.

In order to minimize this influence, the phase lag index\citep{Stam2007c} was proposed as an estimator for the asymmetry of the distribution of phase differences between two time series:
\begin{equation}\label{eq:PLI}
    P = \left| \frac{1}{N} \sum_{j=1}^{N} \sgn \left[\sin \left(\Phi_{a}(j) - \Phi_{b}(j)\right) \right] \right|.
\end{equation}
$P$ is also confined to the interval $[0,1]$, where $P \approx 1$ indicates phase synchronization with finite phase differences.
$P \approx 0$ either indicates no interaction or a distribution of phase differences centered around integer multiples of $\pi$.
The latter may be due to almost identical time series because of a strong interaction or because of the influence of a common source.

More recently, a modification of the phase lag index was proposed which is also resistant to effects of common sources but more robust against noise.
By assigning lower weights to phase differences around 0 and $\pi$, the weighted phase lag index\cite{Vinck2011} is defined as:
\begin{equation}\label{eq:WPLI}
    P_\text{w} = \frac{\left| \sum_{j=1}^{N} \sin \left(\Phi_{a}(j) - \Phi_{b}(j)\right) \right| }{\sum_{j=1}^{N} \left| \sin \left(\Phi_{a}(j) - \Phi_{b}(j) \right) \right|}.
\end{equation}
In case of identical phase time series here we define $P_\text{w}=0$.
$P_\text{w}$ has the same codomain as $P$ and allows to draw the same conclusions about the strength of an interaction.

\subsection*{Generating model time series}
In order to mimic experimental situations and to create a controlled setting for a comparison of the aforementioned estimators, we generated noisy model time series influenced by a common source.
To this end, we used the dynamics of various paradigmatic model systems: coupled phase oscillators as well as coupled structurally identical and non-identical nonlinear oscillators with chaotic dynamics.

First, we consider the Kuramoto model\cite{Acebron2005}, which here consists of two coupled phase oscillators $\Phi_{a}$ and $\Phi_{b}$ with natural frequencies $\omega_a$ and $\omega_b$.
The equations of motion read:
\begin{equation}\label{eq:KO}
    \dot{\Phi}_{a}= \omega_{a} + K_{a} \sin \left( \Phi_{b} - \Phi_{a} \right) + \eta_{a}
\end{equation}
and analogous for $b$.
Here $K_{l}, l \in \{a, b\}$ denotes the coupling strength and $\eta_{l} \in \cal{N}$(0, 0.1) is a small-amplitude noise, which we add to the respective phase dynamics in order to have the oscillators not immediately synchronized for some $K_{l} > 0$.
Choosing initial conditions randomly from $[0, 2\pi)$ and with $\omega_a = 0.8$ and $\omega_b = 1.0$, we integrated Eq.~\ref{eq:KO} using an Euler--Maruyama scheme with a step-size of 0.01 and sampling interval of 0.2.
After discarding $10^5$ transients, we obtained phase time series with 30--40 data points per period and took the sine of the respective phase to generate  observables.

Second, we consider diffusively coupled R\"ossler oscillators\cite{Roessler1976} with a slight mismatch in their natural frequencies $\omega_a = 0.8$ and $\omega_b = 1.0$:
\begin{equation} \begin{split}
    \dot{x}_{a} &= \omega_{a} ~(- y_{a} - z_{a}) + K_{a} (x_{b} - x_{a}), \\
    \dot{y}_{a} &= \omega_{a} ~(x_{a} + 0.15 y_{a}), \\
    \dot{z}_{a} &= \omega_{a} ~(0.2 + z_{a} (x_{a} - 10))
    \label{eq:RO}
\end{split} \end{equation}
and analogous for $b$.
With initial conditions near the attractors Eqs.~\ref{eq:RO} were integrated using the LSODA solver\cite{Hindmarsh1980} with step-size 0.033 and sampling interval 0.198.
As observables we chose the $x$-components after discarding $5 \cdot 10^4$ transients.

\begin{figure*}
\centering
\includegraphics{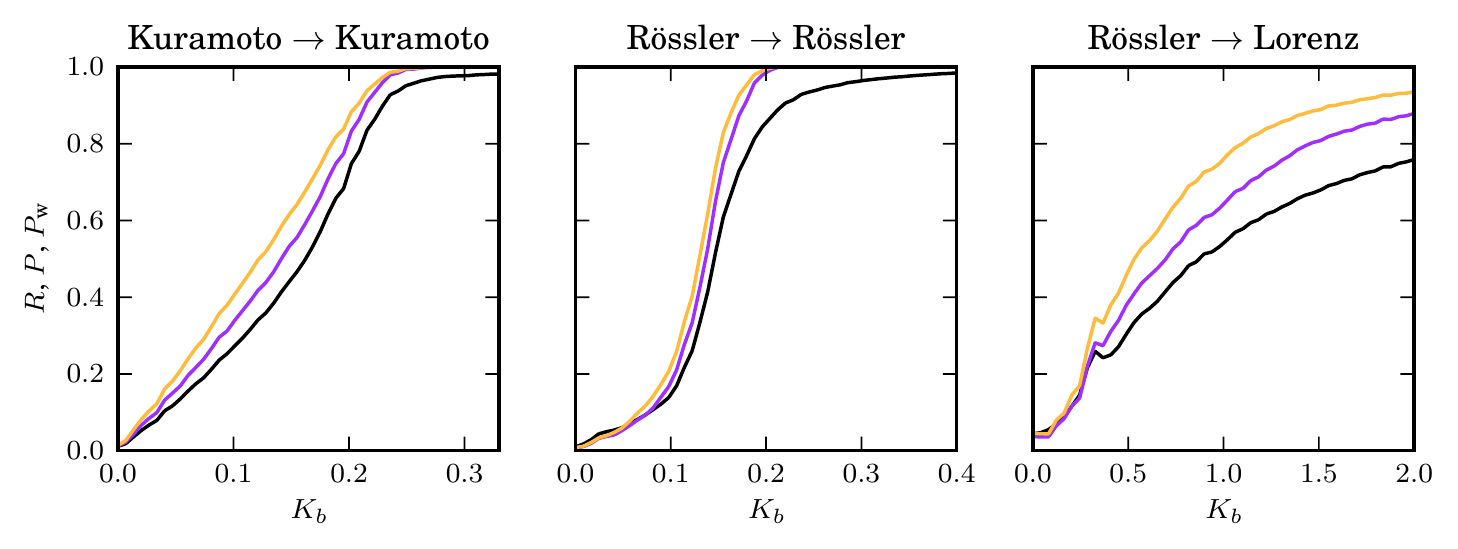}
\caption{
    Dependence of estimators for the strength of interactions on coupling strength.
    Mean values of $R$ (black), $P$ (dark purple), and $P_\text{w}$ (light orange) from 20 realizations of each oscillator system without common source superposition and without noise contamination.}
\label{fig:coup}
\end{figure*}

As a third dynamics, we investigate a R\"ossler oscillator (system $a$) coupled to a Lorenz oscillator\cite{Lorenz1963} (system $b$):
\begin{equation} \begin{split}\label{eq:RLO}
    \dot{x}_{a} &= \omega_{a} ~(- y_{a} - z_{a}) + K_{a} (z_{b} - x_{a} - m), \\
    \dot{y}_{a} &= \omega_{a} ~(x_{a} + 0.15 y_{a}), \\
    \dot{z}_{a} &= \omega_{a} ~(0.2 + z_{a,b} (x_{a} - 10)), \\
    \dot{x}_{b} &= 10 (y_{b} - x_{b}), \\
    \dot{y}_{b} &= 28 x_{b} - y_{b} - x_{b} z_{b}, \\
    \dot{z}_{b} &= (-8/3) z_{b} + x_{b} y_{b} + K_{b} (x_{a} - z_{b} + m),
\end{split} \end{equation}
where $m = 23.5$ compensates the non-vanishing mean of ${z}_{b}$.
If not mentioned otherwise, we set $\omega_{a}=7.5$ to nearly match the speed of the Lorenz oscillator, which in principle allows for phase synchronization.
Initial conditions were near the attractors and integration of Eqs.~\ref{eq:RLO} was carried with LSODA with step-size 0.003 and sampling interval 0.018.
We chose $x_{a}$ and $z_{b}$ as observables after discarding $5 \cdot 10^4$ transients.

Having generated time series $s_{l}(i), l \in \{a, b\}, i = 1, \ldots, N'$ with $N' = 16384$ from observables of the aforementioned coupled oscillators, we next model the influence of a common source.
For this purpose, common source contaminated time series $\tilde s_{l}$ are realized as simple linear superpositions of $s_{l}$, with the exact functional relationships given later.

We then contaminate the time series with different types of measurement noise $\epsilon_{l}, l \in \{a, b\}$ with a noise-to-signal ratio $\nu \in [0, 3.6]$, where $\nu$ is defined as the ratio of the variance of $\epsilon_{l}$ and the variance of $\tilde s_{l}$.
Here we consider $\epsilon_{l}$ to be either Gaussian white noise with zero mean or in-band noise, where the latter is derived from a phase randomized surrogate\cite{Theiler1992} of $\tilde s_{l}$.

Eventually, we derive phase time series $\tilde \Phi_{l}, l \in \{a, b\}$ from $\tilde s_{l}$ via the Hilbert transform.
In order to avoid edge effects, we discarded the first and last 512 instantaneous phase values leaving $N=15360$ data points for the analyses which corresponds to 400--500 oscillations for each of the investigated coupled oscillators.
In the following, we present our findings obtained from 20 realizations of each system for each step of analysis as described above.

\section*{Results from analyses of model data}
\subsection*{Dependence on coupling strength}
First, we investigate the dependence of estimators for the strength of interactions ($R$, $P$, $P_\text{w}$) on the coupling strength for the non-obscured case, i.e., without the influence of common sources and measurement noise.
In Fig.~\ref{fig:coup} we show our findings for the Kuramoto and R\"ossler oscillators, with the slower oscillator ($a$) coupled unidirectionally to the respective faster one with coupling strength $K_b$ ($K_a = 0$), and for the R\"ossler oscillator coupled unidirectionally to the Lorenz oscillator with strength $K_b$ ($K_a = 0$).
For all oscillator systems, estimators increase monotonically with increasing coupling strength $K_b$, but as expected\cite{Kreuz2007} the dependence on $K_b$ is different for the systems.
For the R\"ossler--Lorenz oscillator system, we obtained non-zero values for all estimators already for the uncoupled case ($K_b = 0$) which can be related to the number of data points (indeed, estimators asymptotically approach zero for large $N$).
The dip at $K_b \approx 0.35$ indicates the onset of a deformation of the Lorenz attractor\cite{QuianQuiroga2002}.
For each oscillator system, we observe estimators to exhibit slightly different increases with increasing coupling strength $K_b$, suggesting different routes to complete phase synchronization.
For medium to large values of the coupling strength we have $R \leq P \leq P_\text{w}$.
With $R$, complete synchronization ($R = 1$) is not suggested until much higher coupling strengths than it is for $P$ and $P_\text{w}$, which reflects the respective underlying definitions for complete synchronization\cite{Stam2007c, Vinck2011} (i.e., $\Delta \Phi(t) = \rm const$ and either $\pi > \Delta \Phi(t) > 0$ or $-\pi < \Delta \Phi(t) < 0$, respectively, with $\Delta \Phi(t) \equiv (\Phi_a(t) - \Phi_b(t)) \rm{~mod~} 2 \pi$).

\begin{figure*}
\centering
\includegraphics{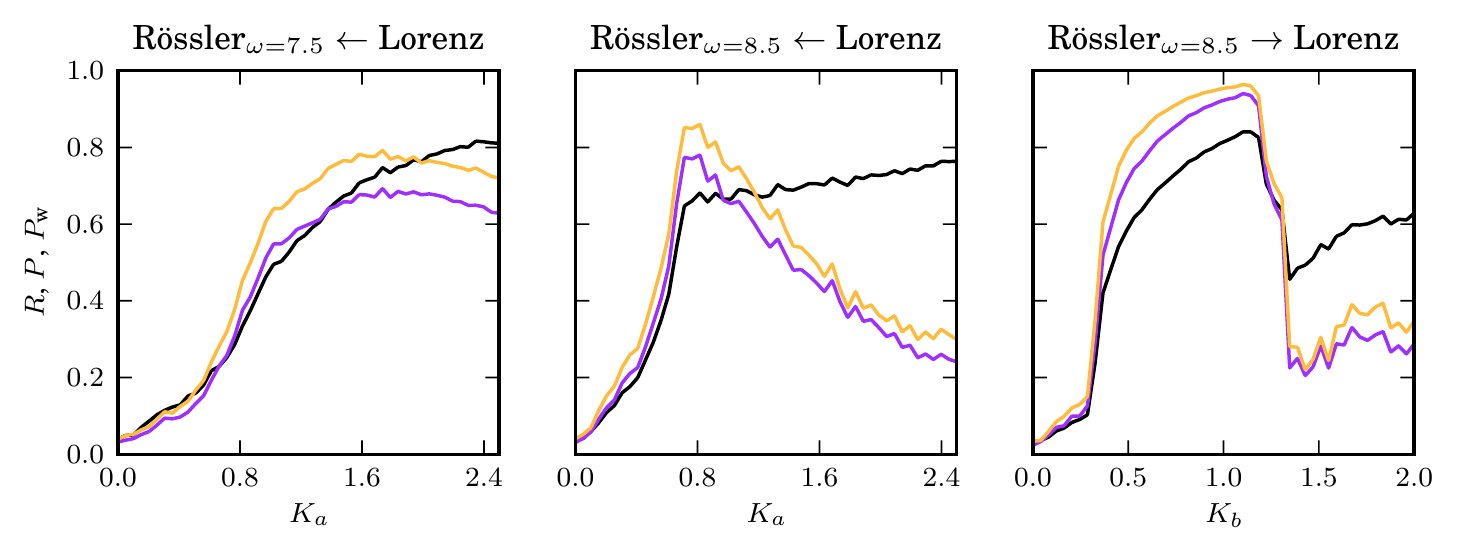}
\caption{
    Same as the right plot of Fig.~\ref{fig:coup} but for reversed coupling (left) and with the R\"ossler oscillator faster than the Lorenz oscillator for both coupling directions (middle and right).}
\label{fig:coup_add}
\end{figure*}

Next we consider the case of a reversed coupling direction and exemplify in Fig.~\ref{fig:coup_add} (left and middle) our findings for the Lorenz oscillator coupled unidirectionally with coupling strength $K_a$ ($K_b = 0$) to the R\"ossler oscillator.
If we choose the eigenfrequency as before ($\omega_a = 7.5$), $R$ exhibits a similar dependence on $K_a$.
Given that estimators are symmetric under exchange of system $a$ and $b$ by definition, this may have been expected.
$P$ and $P_\text{w}$, however, appear to violate this expectation as they decrease for larger coupling strengths, which would even suggest a desynchronization of oscillators.
This effect is even more pronounced if we choose the R\"ossler oscillator to be faster than the Lorenz oscillator by setting $\omega_a = 8.5$ (Fig.~\ref{fig:coup_add} middle), and can also be observed for all estimators if we reverse the coupling direction for these oscillators (Fig.~\ref{fig:coup_add} right).

These findings already indicate that, even under ideal conditions, care must be taken when inferring the strength of interactions and/or different phase synchronization regimes with the various estimators.
A direct comparison between their values is difficult and might lead to misinterpretations, particularly when investigating field data.

\subsection*{Influence of common sources}

\begin{figure}
\centering
\includegraphics{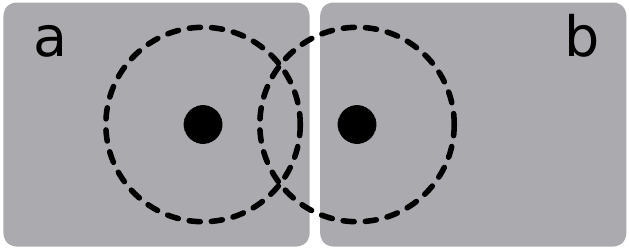}
\caption{
    Schematic of common source contamination: the dynamics of systems $a$ and $b$ are measured via some sensors (black dots) that may pick up not only the respective dynamics but also a mixture, depending on various factors such as placement or pick-up range (dashed lines).}
\label{fig:cs_scheme}
\end{figure}

In the following, we consider various schemes for common source contamination that mimic typical experimental situations.
To this end, let us assume $s_{l}, l \in \{a, b\}$ to represent time series of observables of spatially extended dynamical systems $a$ and $b$.
These time series are to be measured via two or more sensors and from these measurements characteristics of a possible interaction are to be derived.
Due to a lack of detailed knowledge about the systems as well as to other limitations (such as size of sensors or a limited spatial sampling), the non-ideal measurement (here with two sensors) will result in time series $\tilde s_{l}$ representing some superposition of $s_{l}$, i.e., they are contaminated---to a varying degree---by a common source (Fig. \ref{fig:cs_scheme}).

With our first scheme, we consider the case that only one time series is contaminated with different amounts of a common source, while the other is unaffected.
This would represent the condition of having placed one sensor ideally (capturing the dynamics of system $a$ only) and the other sensor such that it captures the dynamics of systems $b$ and to varying degree also that of system $a$, or vice versa:
\begin{equation}\begin{split}
    \tilde s_a(j) &= (1 - \alpha) s_a(j) + \alpha s_b(j),~\tilde s_b(j) = s_b(j),~\text{or}\\
    \tilde s_b(j) &= (1 - \alpha) s_b(j) + \alpha s_a(j),~\tilde s_a(j) = s_a(j),
    \label{eq:cs_single}
\end{split}\end{equation}
where $\alpha \in [0, 1)$ controls the amount of superposition.

With our second scheme, we consider the case that both time series are contaminated with different amounts of a common source.
This would represent the condition of having placed both sensors non-optimally such that each captures a mixture of the dynamics of both systems:
\begin{equation}\begin{split}
    \tilde s_a(j) &= (1 - \alpha) s_a(j) + \alpha s_b(j),\\
    \tilde s_b(j) &= (1 - \alpha) s_b(j) + \alpha s_a(j),
    \label{eq:cs_both}
\end{split}\end{equation}
where $\alpha \in [0, 0.5)$ controls the amount of mixing.

\begin{figure*}
\centering
\includegraphics{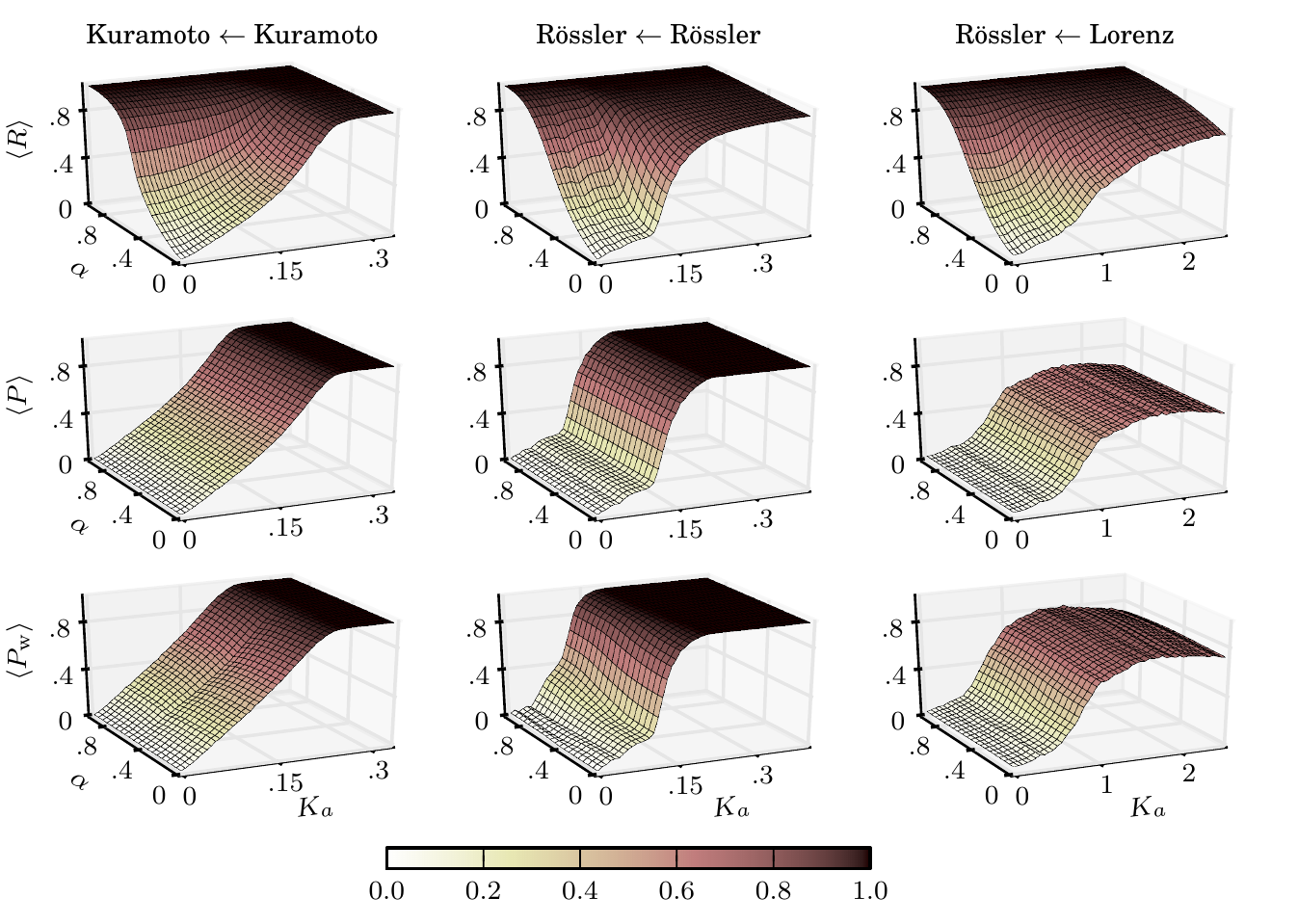}
\caption{
    Dependence of estimators for the strength of interactions on coupling strength
    $K_a$ and on the amount of common source superposition $\alpha$.
    Only time series from responding oscillators were influenced by common sources (cf. Eq.~\ref{eq:cs_single}).
    Brackets denote mean values of estimators from 20 realizations of each oscillator system.}
\label{fig:cs}
\end{figure*}

With our third scheme, we consider the case that the time series $\tilde s_{a}$ and $\tilde s_{b}$ are unaffected, and, lacking detailed knowledge about systems $a$ and $b$, we have measured an additional time series $\tilde s_{c}$ that constitutes a mixture of their dynamics.
This would represent the condition of having placed sensors for systems $a$ and $b$ optimally and the other sensor such that it over-samples their dynamics:
\begin{equation} \begin{split}
    \tilde s_a = s_a,~\tilde s_b = s_b,~\tilde s_c = \alpha s_a + (1 - \alpha) s_b,
\label{eq:cs_oversampl}
\end{split}\end{equation}
where $\alpha\in [0, 1)$ controls the amount of mixing.
For this case, we estimate the interaction from the pairs $(\tilde s_a, \tilde s_b)$, $(\tilde s_a, \tilde s_c)$, and $(\tilde s_b, \tilde s_c)$.

Using these contamination schemes, we investigated the dependence of the estimators on $\alpha$ and on the coupling strength $K_{l}, l \in \{a, b\}$, taking also into account the additional dependencies identified above (exchange of driver and responder, choice of eigenfrequencies).

As an example, we show in Fig.~\ref{fig:cs} the dependencies of estimators on the coupling strength $K_a$ and on the amount of common source superposition $\alpha$ for the case that the faster oscillator (driver) is coupled unidirectionally to the slower one (responder) and that only the time series from the responding oscillator is contaminated by a common source (cf. Eq.~\ref{eq:cs_single}).
We observe $R$ to be strongly influenced by common sources: $R$ increases with increasing values of $\alpha$ even for uncoupled oscillators, which would misleadingly indicate a coupling between them.
In contrast, $P$ is not affected by common sources for any oscillator system.
We note that other coupling and contamination schemes (we will discuss the third contamination scheme separately) yielded very similar dependencies for both $R$ and $P$.
In general, $P_\text{w}$ exhibits the same dependence on $\alpha$ as $P$, but we observe additional influences of common sources on this estimator that depend on the investigated oscillator system.
For the Kuramoto oscillators and for the first contamination scheme, we observe $P_\text{w}$ to slightly decrease around $\alpha = 0.5$ for a wide range of coupling strengths $K_a$.
This effect appears to be specific for this oscillator system, as it can be observed around this same $\alpha$ for all investigated coupling and contamination schemes, and it does not depend sensitively on the choice of internal parameters (natural frequencies, dynamic noise).
For the coupled R\"ossler oscillators and R\"ossler-Lorenz oscillator system results are highly inconsistent and $P_\text{w}$ either increases or decreases with increasing $\alpha$ in different coupling regimes depending on which coupling or contamination scheme is used (data not shown).
Here we observed deviations from the ideal case of up to 0.45.

For our third contamination scheme (cf. Eq.~\ref{eq:cs_oversampl}), results for each pairwise analysis are, by construction, identical to those obtained for the non-obscured case with the time series pair $(\tilde s_a, \tilde s_b)$ and for the first contamination scheme with time series pairs $(\tilde s_a, \tilde s_c)$ and $(\tilde s_b, \tilde s_c)$.
We will nevertheless briefly present them again in order to illustrate problems that may occur when analyzing the dynamics of spatially over-sampled systems subject to common source contaminations.
With $(\tilde s_a, \tilde s_b)$, all estimators well reflect the underlying coupling.
With the other two pairs and for $\alpha > 0$, $R$ clearly overestimates the strength of interaction (even for uncoupled systems), which would erroneously indicate three interacting systems.
For $0 < \alpha < 1$, both $P$ and $P_\text{w}$
(apart from the above mentioned shortcomings for this estimator)
estimate the strengths of interaction from the three time series pairs to be identical, making it difficult to distinguish between three (non-)interacting and two over-sampled systems.

\subsection*{Impact of common sources and of measurement noise}

\begin{figure*}
\centering
\includegraphics{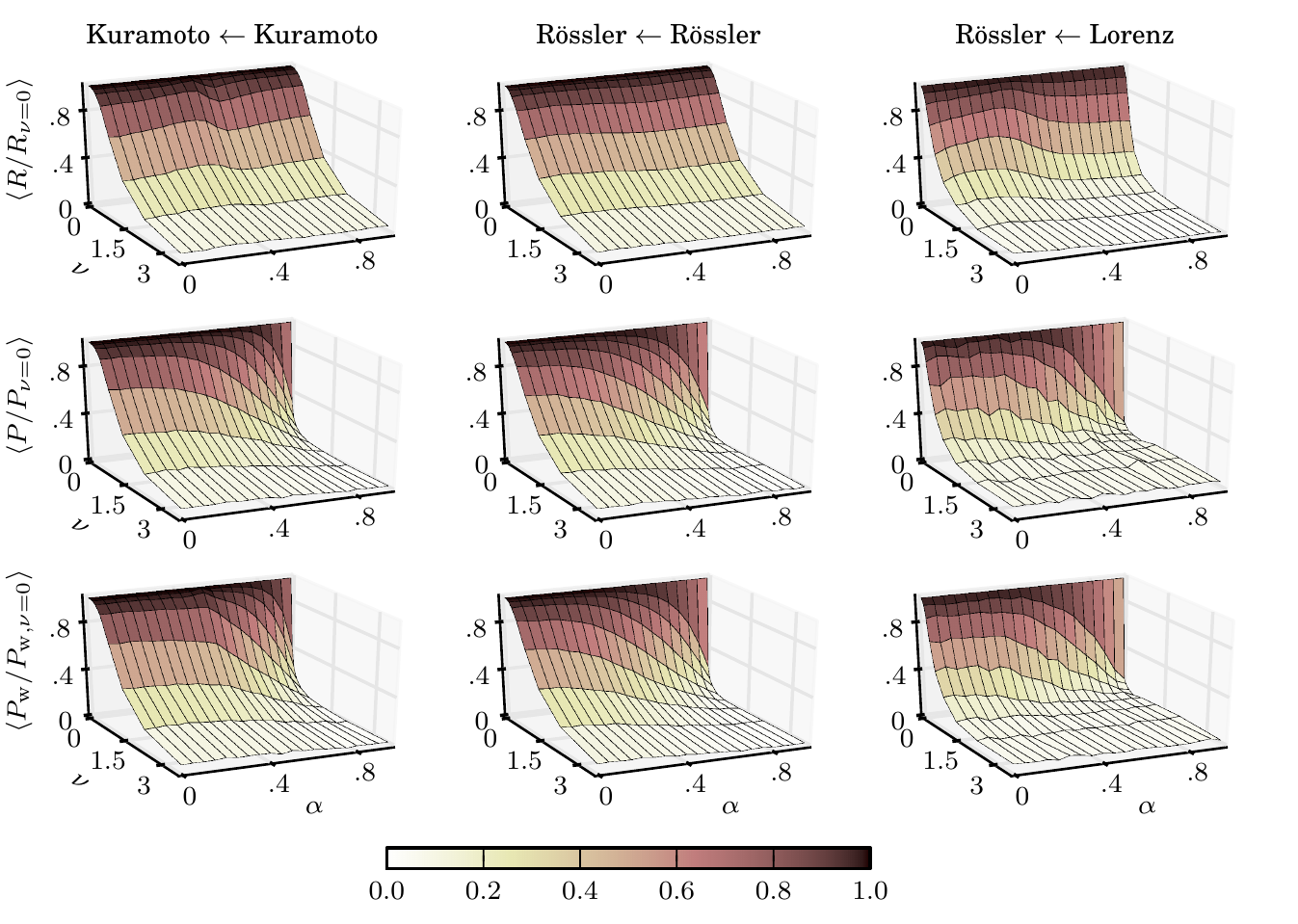}
\caption{
    Dependence of estimators for the strength of interactions on noise-to-signal ratio $\nu$ and on amount of common source superposition $\alpha$ for exemplary coupling strengths: $K_a=0.135$ (Kuramoto oscillators), $K_a=0.122$ (R\"ossler oscillators) and $K_a=0.561$ (R\"ossler-Lorenz oscillators); cf. Figs.~\ref{fig:coup} and~\ref{fig:coup_add}.
    Time series were contaminated with Gaussian white noise, but only time series from responding oscillators were influenced by common sources.
    In each system, the faster oscillator is coupled unidirectionally to the slower one.
    For each $\alpha$, mean relative values of estimators from 20 realizations of each oscillator system are shown, normalized to the respective values of the noise-free cases.}
\label{fig:noise_gauss}
\end{figure*}

We now investigate the impact of different types of measurement noise together with that of common sources.
For a contamination with Gaussian white noise and with common sources according to Eq.~\ref{eq:cs_single}, we show in Fig.~\ref{fig:noise_gauss}
normalized estimator values depending on the noise-to-signal ratio $\nu$ and on the amount of common source superposition $\alpha$ for the oscillator systems using some intermediate coupling strength.
Findings were similar for other coupling schemes and with common source contaminations according to Eq.~\ref{eq:cs_both}.
For all $\alpha$, estimator values decrease with increasing $\nu$, but we observe qualitative and quantitative differences.
For $\alpha = 0$, $P$ and $P_\text{w}$ decline similarly as $R$.
For strong common source contaminations, however, values of both $P$ and $P_\text{w}$ decrease more rapidly, and for $\nu \approx 1$ both estimators loose their ability to characterize the strength of interaction.

\begin{figure*}
\centering
\includegraphics{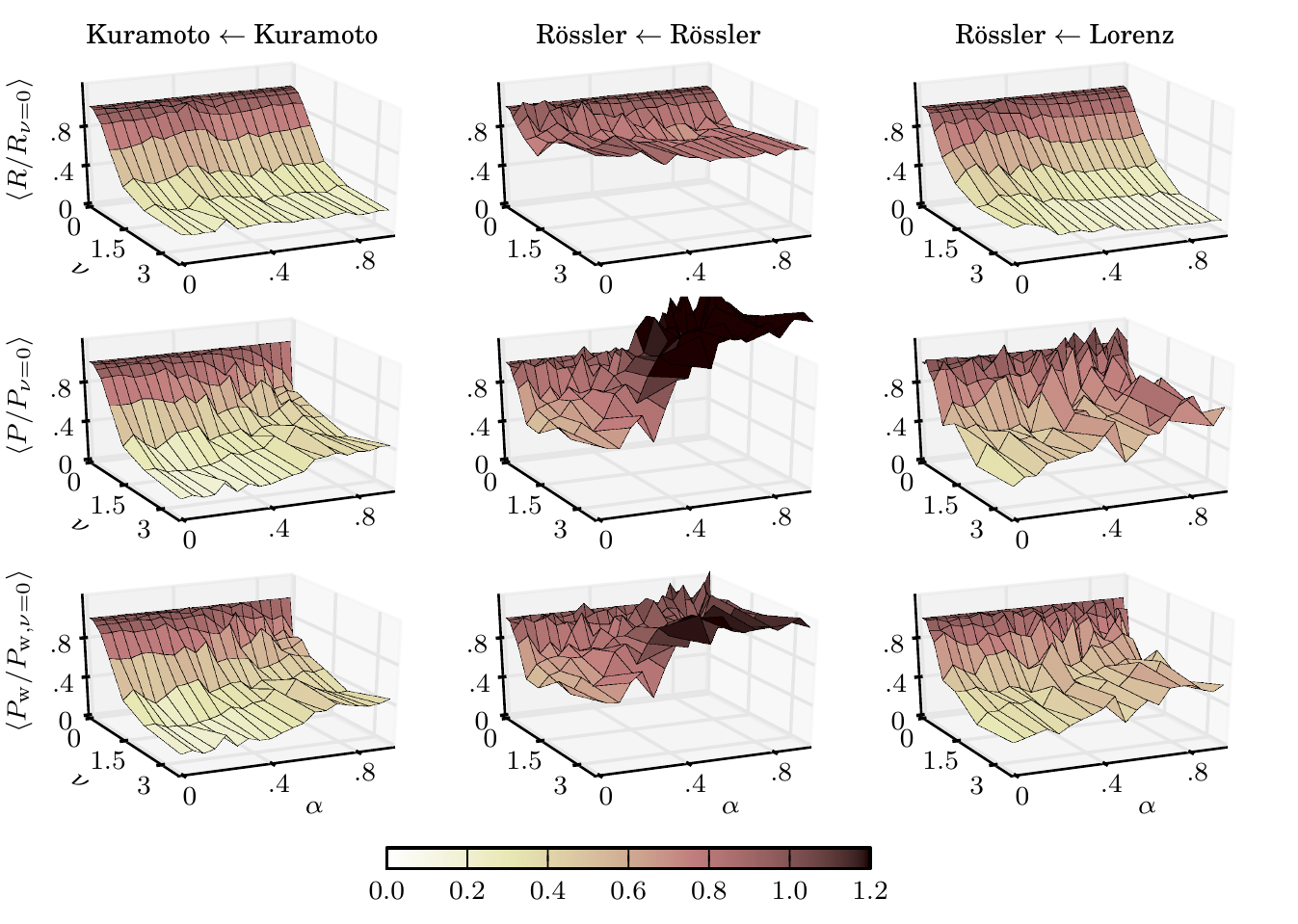}
\caption{
    Same as Fig.~\ref{fig:noise_gauss} but for in-band noise.}
\label{fig:noise_iso}
\end{figure*}

If time series are contaminated with in-band noise, estimators exhibit a highly inconsistent dependence on the noise-to-signal ratio $\nu$ and on the amount of common source superposition~$\alpha$ (Fig.~\ref{fig:noise_iso}).
For the Kuramoto oscillators, estimator values decrease with increasing $\nu$ for all values of $\alpha$.
When compared to the contamination with Gaussian white noise, the decline is less steep and all estimators exhibit a higher variability.
For the other oscillator systems, the dependence of estimators on $\nu$ and $\alpha$ appears to be additionally influenced by the spectral contents of the respective dynamics, which for some $\nu$ and $\alpha$ can even result in estimator values exceeding those for the noise-free case.

When comparing the dependencies of $P$ on $\alpha$ and $\nu$ for all types of contaminations with noise and common sources with the dependencies of
$P_\text{w}$, we observe some significant differences (see, e.g, data for the R\"ossler-Lorenz oscillator system for small values of $\nu$ in Fig.~\ref{fig:noise_gauss}) that would underline the improved robustness against noise of the latter estimator\cite{Vinck2011}.
Our findings are, however, not consistent as they strongly depend on the investigated dynamics as well as on coupling schemes and on the scheme used for common source contamination.
Only for the Kuramoto oscillators can we observe $P_\text{w}$ to decrease more slowly with increasing noise-to-signal ratio $\nu$ than $P$ in all investigated cases.
For the other investigated oscillator systems, significant differences between estimator performances are scarce, if any.

Before closing this section, we briefly summarize our findings obtained from numerically simulating typical experimental situations.
When time series of observables of (non\nobreakdash-)interacting systems are contaminated with common sources, the strength of interaction is over-estimated when using the mean phase coherence~$R$.
In contrast, the phase lag index~$P$ and to a lesser extent also the weighted phase lag index~$P_\text{w}$ are unaffected by common sources.
Ambiguities may occur, however, with all estimators in case of spatially over-sampled systems.
When time series are strongly contaminated with noise, all estimators loose their ability to characterize the strength of interaction, as expected.
However, this already happens for smaller noise levels when common sources are present.

\section*{Measuring the strength of interactions in the human epileptic brain}
Phase-based methods for the detection of the strength of interactions between different brain regions using electroencephalographic (EEG) time series have been repeatedly demonstrated to yield meaningful results for such different problems as  the identification of the seizure-generating area of the brain (epileptic focus)\cite{Bialonski2006a, Osterhage2007, Schevon2007, Zaveri2009, Warren2010, Ortega2011}, the detection of precursors of epileptic seizures\cite{Mormann2000, Mormann2003a, LeVanQuyen2005, Mormann2005, Kuhlmann2010}, or the advancement of our understanding about complex synchronization phenomena underlying seizure dynamics\cite{Netoff2002, Dominguez2005, Schindler2007b, Kiss2008, Serletis2013}, mental disorders\cite{Bob2008}, cognition\cite{Fell2001, Canolty2006, Mormann2008b, Palva2010b}, and sleep\cite{Mezeiova2012, Moroni2012}.
Nevertheless, the interpretability of findings might be limited due to potentially confounding variables, including those that are particularly related to the EEG recording:
Closely spaced sensors are very likely to pick up the dynamics of the same (common) sources, and the necessity to choose an active reference sensor, which is a notoriously ill-defined problem\cite{Yao2005}.
Both factors are known to affect phase-based estimators for the strength of interactions\cite{Guevara2005, Tognoli2009, Hu2010}.

Complementing our numerical studies on the relative merit of estimators for the strength of interactions ($R$, $P$, $P_\text{w}$), we investigated consistency and temporal variations of the strength of interactions within and between brain regions using long-lasting, multichannel intracranial EEG recordings from an epilepsy patient undergoing presurgical evaluation.
The referential recording with sensors that sample directly---due to clinical reasons---various brain regions with different spatial resolutions allows us to analyze the impact of the aforementioned influencing factors in realistic situations.

\begin{figure*}
\centering
\includegraphics[width=1.0\textwidth]{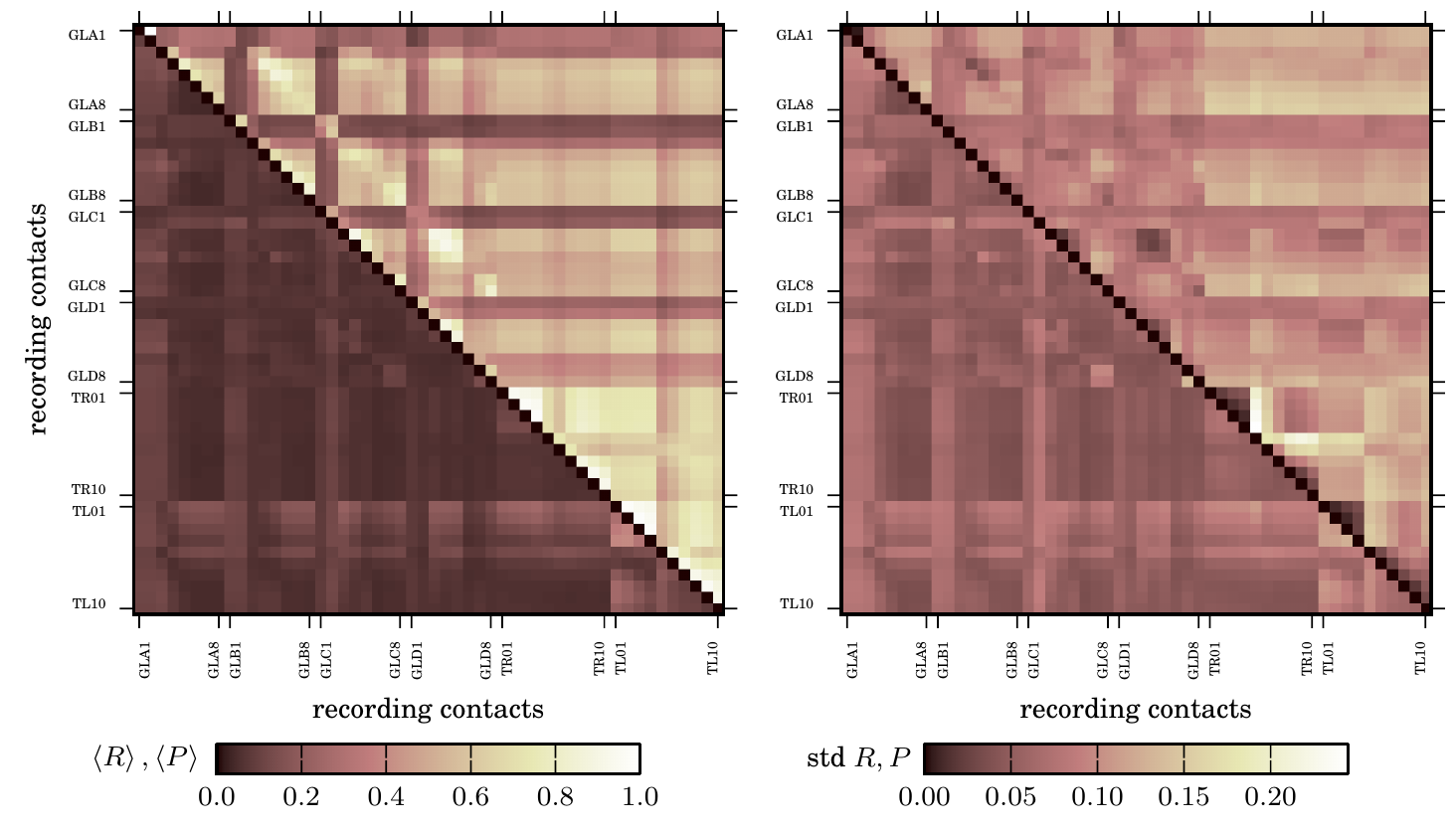}
\caption{
    Mean (left) and standard deviation (right) of the strength of interactions between EEG time series from all pairs of recording contacts (Fig. \ref{fig:impscheme}) estimated with $R$ (upper triangle) and with $P$ (lower triangle) from a recording of \unit[20]{h} duration.
    Contacts GLA1 and GLA2 were used as recording reference, contacts GLD3 and GLD4 covered a structural lesion, and the presurgical workup identified contact GLD6 to cover the epileptic focus (seizure-generating brain region).}
\label{fig:field_est}
\end{figure*}

\begin{figure}
\centering
\includegraphics[width = 0.45\textwidth]{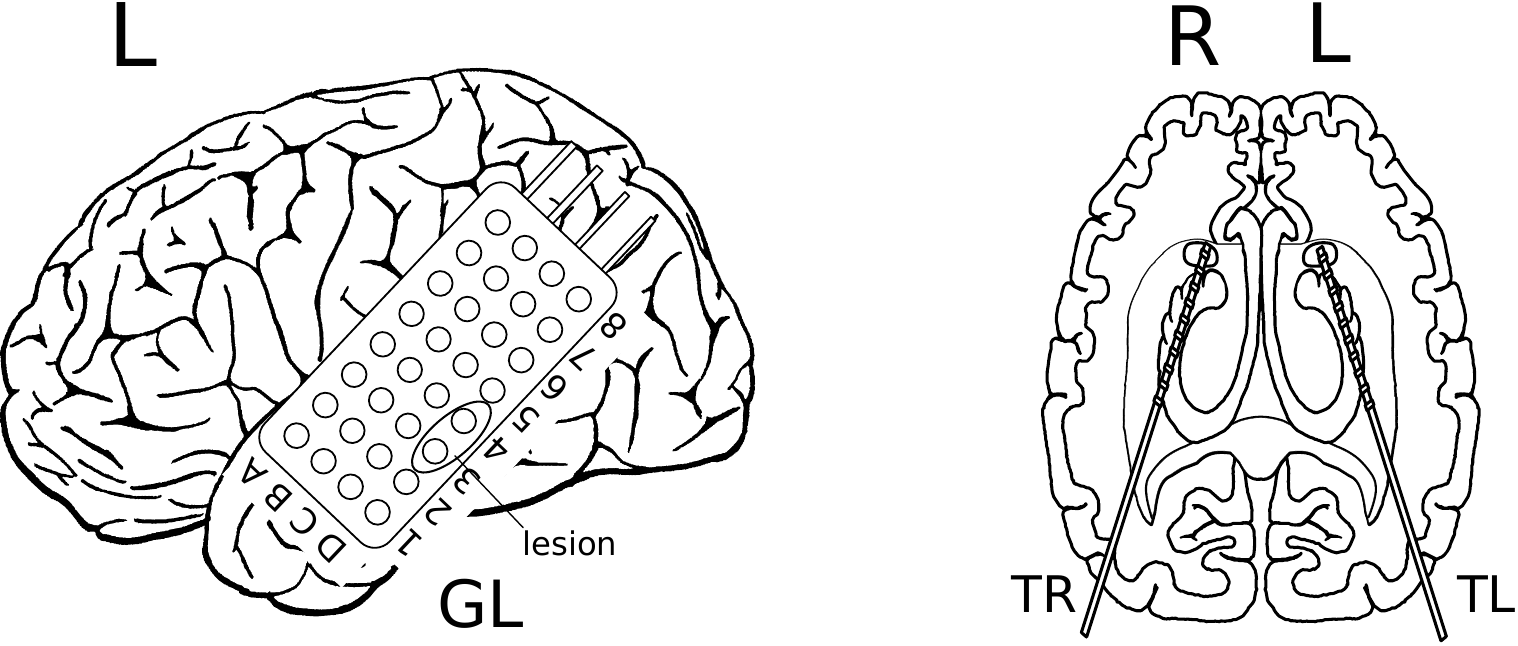}
\caption{
    Schematics of the electrode grid placed over the temporal lateral neocortex (left) and of bilateral intrahippocampal depth electrodes (right; axial view).}
\label{fig:impscheme}
\end{figure}

The patient had signed informed consent that her/his clinical data might be used and published for research purposes, and the study protocol had previously been approved by the local medical ethics committee.
The EEG was recorded prior to surgery via two intrahippocampal depth electrodes (each equipped with 10 cylindrical contacts of length \unit[2.5]{mm} and an intercontact distance of \unit[4]{mm}; implanted stereotactically in the medial temporal lobes) and from grid electrodes (rectangular flexible grid of $8 \times 4$ contacts with an intercontact distance of \unit[10]{mm}; placed subdurally onto the temporal lateral neocortex) referenced against the average activity of two recording contacts (GLA1 and GLA2) distant to the seizure-generating brain region (GLD6, Fig. \ref{fig:impscheme}).
Data were band-pass filtered between \unit[0.1--70]{Hz} and sampled at \unit[200]{Hz} using a \unit[16]{bit} analog-to-digital converter.

Here we consider a continuous recording of \unit[20]{h} duration that started in the morning and covered different physiologic and pathophysiologic, disease-related states of the patient; no seizures occurred during this recording.
Prior to estimating strengths of interactions with $R$, $P$, and $P_\text{w}$ in a time-resolved manner, we digitally band-pass filtered the data between \unit[1--45]{Hz} (2nd order Butterworth characteristic), suppressed possible contributions of the power line frequency using a notch filter, and derived phase time series via the Hilbert transform\cite{Boashash1992_book, Mormann2000}.

As a compromise between the statistical accuracy for the calculation of estimators and approximate stationarity, we divided the data into non-overlapping segments of \unit[20.48]{s} duration (corresponding to 4096 data points).
This allowed us to calculate the estimators for each combination of pairs of recording contacts in a moving-window fashion.

First, we concentrate on the spatial distribution of interaction strengths and check whether consistently interacting brain regions can be identified.
To this end, we performed a time-averaging over all windows, resulting in a matrix of averaged strengths of interactions for each estimator ($R$, $P$, $P_\text{w}$).
In Fig.~\ref{fig:field_est}, we show averages and standard deviations for $R$ and $P$; findings for $P_\text{w}$ closely resemble those for $P$ and are not shown.

With $R$, we observe a quite distinct spatial distribution of interaction strengths:
Depending on brain regions, highest values are mostly confined to contact pairs that are nearest neighbors, and medium to low values roughly reflect long-ranged interactions.
The hippocampal formations (sampled with depth electrodes TL and TR) exhibit the highest intra- and interhemispheric interaction strengths and we can identify the known functionally definable subregions for these brain structures\cite{Mormann2008b}.
Highest interaction strengths in the temporal lateral neocortex are confined to contact pairs covering the lesion (contacts GLD3 and GLD4) and brain tissue surrounding the lesion (contacts GLC3, GLC4, and GLD5).
In line with previous studies\cite{Schevon2007, Warren2010} we observe the seizure-generating brain area (contact GLD6) to exhibit only moderately increased interaction strengths with other brain areas.
Even lower values can be observed for interactions that comprise the reference contacts (GLA1 and GLA2) and their neighboring contacts with all other sampled brain regions.

With $P$, we observe estimated interaction strengths to be quite low overall, including several nearest neighbor contact pairs that exhibit medium to high values with $R$, which would generally point to the presence of strong influences of common sources.
There are, however, exceptions to this rule, as can be seen, e.g., for pairs comprising contacts TL01 to TL04 that densely sample the dynamics of a circumscribed brain region.
Estimated interactions within this region are strongest, and it seems to be interacting with almost all sampled brain regions throughout the recording.
There is a number of other contact pairs (comprising TR01 to TR04, GLB5, GLC3, GLC5, GLC7, GLC8, GLD4, GLC7,and GLC8) that also exhibit slightly increased interaction strengths.
Some of these contacts sample the activity of brain areas surrounding the structural lesion and the seizure-generating region.
Unexpectedly, we observe medium values for strengths of interactions between brain regions sampled by the reference contacts GLA1 and GLA2 (and to a lesser extent also for neighboring contacts CLA3, CLB1, CLB2, CLC1, and CLC2) and almost all other brain regions.
In general, known and consistent spatial interactions can be observed with $P$, if at all, for some brain regions only.

When comparing the spatial distributions of means and standard deviations of interaction strengths (Fig.~\ref{fig:field_est}), we observe that  high mean values of $R$ are associated with low values of the standard deviations for most contact pairs. For $P$, however, high mean values are almost always associated with high values of the standard deviations.
In the following we therefore concentrate on the temporal variability of interaction strengths.
As above, findings for $P_\text{w}$ closely resemble those for $P$ and are not shown.

\begin{figure*}
\centering
\includegraphics[width=0.95\textwidth]{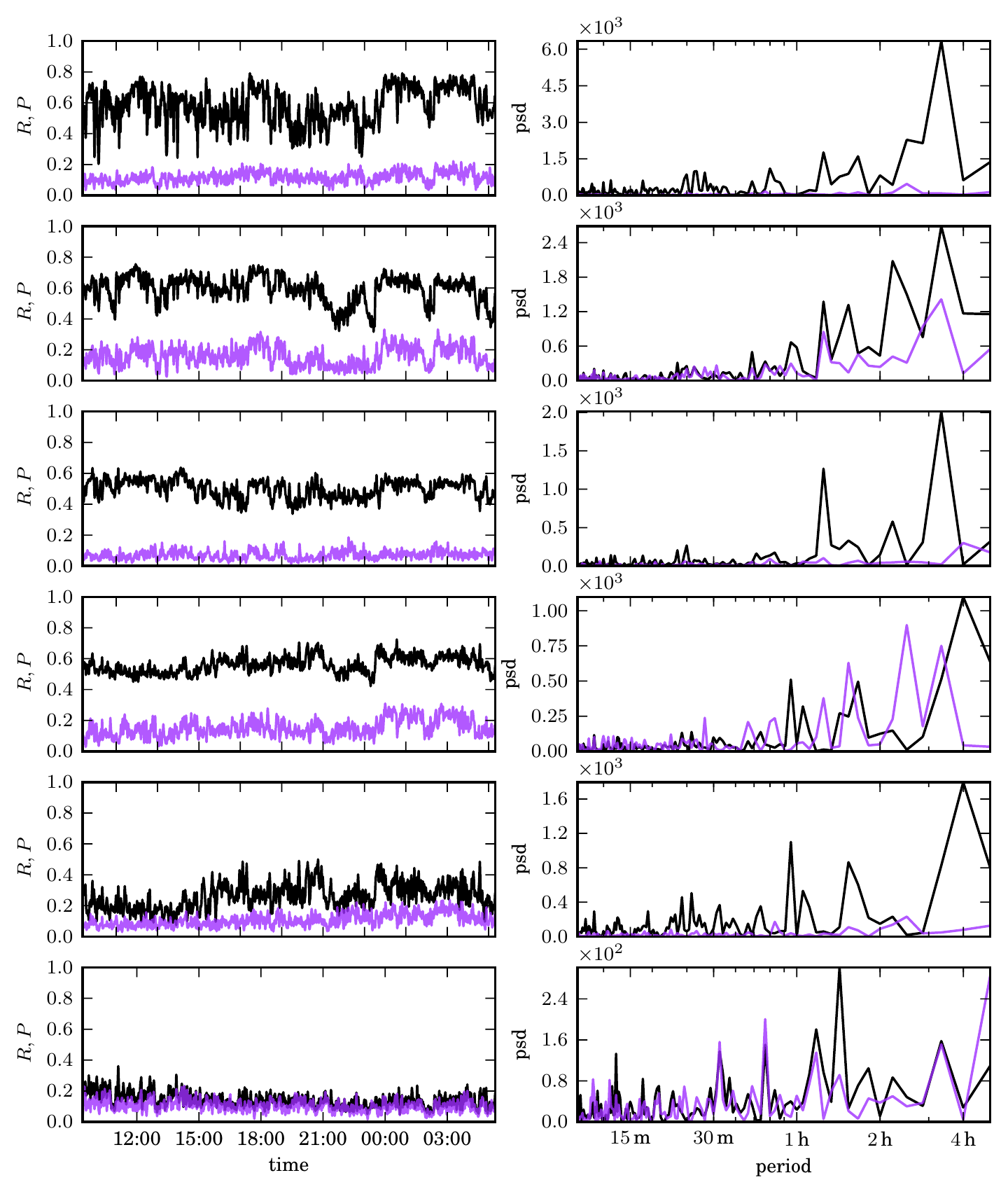}
\caption{
    Left: exemplary temporal evolutions of interaction strengths estimated with $R$ (black) and with $P$ (purple) for short- and long-ranged interactions (top to bottom: contact pairs (TR06, TL03), (TR06, TL05), (GLB4, GLB3), (GLB4, GLA3), (TR07, GLA3), and (TR02, GLB1); Fig. \ref{fig:impscheme}).
    Recording time was \unit[20]{h}.
    For readability, time series are smoothed using a moving Hamming window over 10 data points (corresponding to \unit[4.96]{min}).
    Right: power spectral density estimates of unsmoothed time series of $R$ and $P$.}
\label{fig:field_detail}
\end{figure*}

In rows 1 and 2 of Fig.~\ref{fig:field_detail}, we show temporal evolutions of $R$ and $P$ together with their respective estimates of the power spectral density\cite{Press2007} for selected interactions between the right (contact TR06) and left (contacts TL03 and TL05) hippocampal formation.
With $R$, we observe large fluctuations over time with some temporal structure in the data which seems to be partly periodic.
The power estimates of the spectral density indicate strong contributions from processes acting on timescales of approximately \unit[2--4]{h}, which can probably be related to sleep architecture.
With $P$, a similar patterning can be observed only for the data from contact pairs (TR06, TL05) but not for the data from the other pair (with TL03 being \unit[8]{mm} apart from TL05), despite a comparable average strength of interactions.
Similar findings can be obtained for short-ranged (rows 3 and 4 of Fig.~\ref{fig:field_detail}) as well as for long-ranged interactions (rows 5 and 6 of Fig.~\ref{fig:field_detail}), and independent on the average level of the strength of interactions.

\begin{figure}
\centering
\includegraphics[width = 0.5\textwidth]{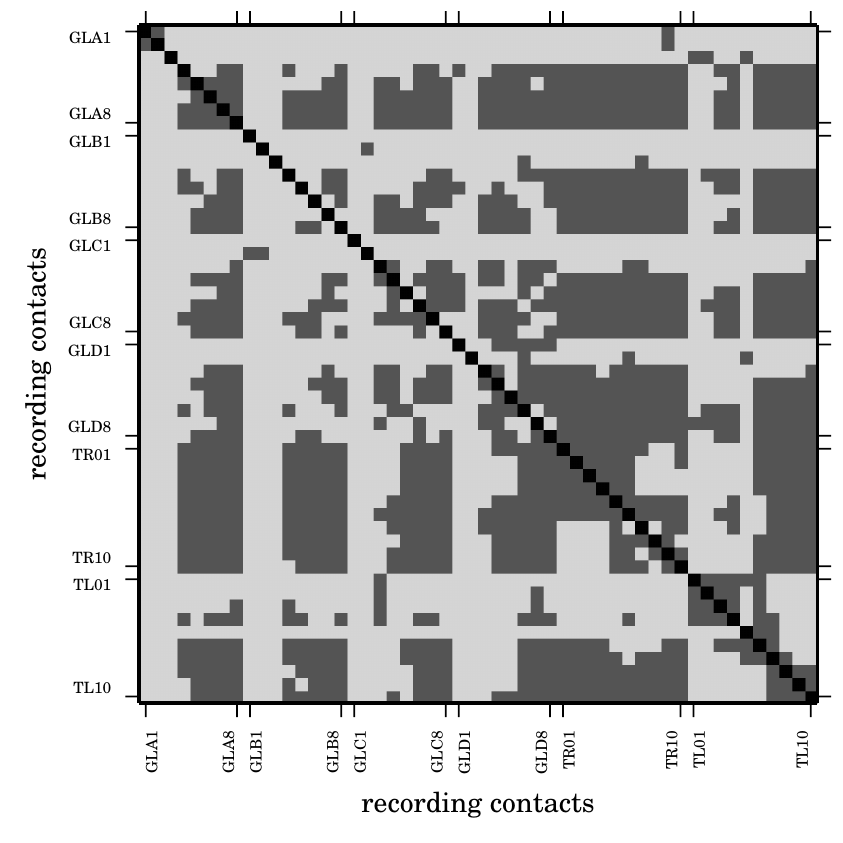}
\caption{
    Equivalence (light gray) and non-equivalence (dark gray) of power spectra of temporal evolutions of $R$ and $P$ (upper triangle) and of $R$ and $P_\text{w}$ (lower triangle) for each combination of pairs of recording contacts. Recording time was \unit[20]{h} and statistical equivalence of power spectra (significance level $p \leq 0.05)$ was tested for all frequencies up to the Nyquist frequency.}
\label{fig:field_pow_eq}
\end{figure}

For completeness, we show in Fig.~\ref{fig:field_pow_eq} results obtained from testing for the statistical equivalence of power spectra\cite{Bendat2000} of the temporal evolutions of $R$ and $P$ and of $R$ and $P_\text{w}$ for each combination of pairs of recording contacts.
At first glance, the spatial patterning of equivalence (non-equivalence) roughly resembles the patterning of high (low) average values of estimators for the strength of interactions (see Fig.~\ref{fig:field_est}): high average values of $P$ (or $P_\text{w}$) are quite often associated with equivalent power spectra, which would point to an issue of the resolution of estimators when approaching the lower (upper) bound of their codomain.
As expected, there are exceptions to this rule, most notably for long- and short-ranged interactions involving contacts sampling the hippocampal formations as well as brain areas surrounding the structural lesion and the seizure-generating region, as already reported on above.

Before closing this section, we briefly summarize our findings obtained from analyzing the strengths of interactions in the epileptic brain using long-lasting, multichannel intracranial EEG recordings.
Due to the dense spatial sampling and due to the necessity to choose an active reference sensor, we expected the data to be influenced by common sources.
With the mean phase coherence~$R$, we could confirm well-known findings concerning the spatial distribution of average strengths of short- and long-ranged interactions as well as their modulation due to various physiologic and pathophysiologic processes that act on different timescales.
Findings obtained with the phase lag index~$P$ appear to suggest strong influences of common sources.
They raise, however, doubts whether the observed discrepancies (also in comparison with the findings obtained with $R$, provided they allow a reasonable interpretation) can fully be explained by the influence of common sources.
It remains to be shown whether a reduction of the influence of common sources leads to a loss of important spatial and temporal aspects of the interaction dynamics.
With our applications we could not follow the claimed advantages of the weighted phase lag index~$P_\text{w}$ over $P$.

\section*{Conclusion}
We investigated the relative merit of a widely used phase-based estimator for the strength of interaction (mean phase coherence~$R$) together with extensions (phase lag index~$P$ and weighted phase lag index~$P_\text{w}$) that had been designed to be immune to common sources and more robust against noise.
Using the dynamics of paradigmatic model systems, we generated time series subjected to various influencing factors, at the same time mimicking some typical experimental situations.
Eventually, we investigated---in a time-resolved manner---the strength of interactions between various brain regions from an epilepsy patient, using long-lasting, multi-channel, invasive electroencephalographic recordings.

With our simulation studies, we could, in general, confirm the advantages of the improved estimators $P$ and $P_\text{w}$ over $R$.
Nevertheless, we also identified cases that can lead to ambiguities with all estimators, namely when spatially oversampling interacting systems.
One is quite often confronted with such a situation, particularly when investigating systems with only poorly understood dynamics.

Our findings obtained from analyzing the interaction dynamics of various brain regions indicate that spurious correlations in the data, which can be regarded as being induced by the recording procedure, can be reduced with $P$ and $P_\text{w}$.
Important and well-known spatial and temporal aspects of the interaction dynamics, however, appear to be lost with these estimators, which calls for further investigations on their effectiveness.
Without more detailed information, a direct comparison of values obtained with the different estimators is discouraged, although all of them are based on the concept of phase synchronization.
Apart from the influence of common sources, there are other confounding variables that had been identified to affect estimates for the strength of interaction.
Here we mention indirect interactions that can be differentiated from direct interactions with multivariate phase-based methods using partialization analysis\cite{Schelter2006b, Kralemann2013, Kralemann2014}.
Nevertheless, methods that allow one to effectively reduce the influence of these and other confounding variables are still missing.
It remains to be shown whether recently proposed methods for an improved phase extraction\cite{Kralemann2011, Schwabedal2012} or specifically designed surrogate techniques can be of help to better delineate functional from spurious interactions between dynamical systems.

\section*{Acknowledgements}
The authors are grateful to G. Ansmann, S. Bialonski and S. Werner for critical comments on earlier versions of the manuscript.
This work was supported by the Deutsche Forschungsgemeinschaft (Grant No. LE660/5\nobreakdash-2).

\end{document}